\PassOptionsToPackage{unicode}{hyperref}
\PassOptionsToPackage{hyphens}{url}
\PassOptionsToPackage{dvipsnames,svgnames,x11names}{xcolor}
\documentclass[
  11pt,
]{article}
\usepackage{amsmath,amssymb}
\usepackage{iftex}
\ifPDFTeX
  \usepackage[T1]{fontenc}
  \usepackage[utf8]{inputenc}
  \usepackage{textcomp} % provide euro and other symbols
\else % if luatex or xetex
  \usepackage{unicode-math} % this also loads fontspec
  \defaultfontfeatures{Scale=MatchLowercase}
  \defaultfontfeatures[\rmfamily]{Ligatures=TeX,Scale=1}
\fi
\usepackage[]{mathptmx}
\ifPDFTeX\else
\fi
\IfFileExists{upquote.sty}{\usepackage{upquote}}{}
\IfFileExists{microtype.sty}{% use microtype if available
  \usepackage[]{microtype}
  \UseMicrotypeSet[protrusion]{basicmath} % disable protrusion for tt fonts
}{}
\makeatletter
\@ifundefined{KOMAClassName}{% if non-KOMA class
  \IfFileExists{parskip.sty}{%
    \usepackage{parskip}
  }{% else
    \setlength{\parindent}{0pt}
    \setlength{\parskip}{6pt plus 2pt minus 1pt}}
}{% if KOMA class
  \KOMAoptions{parskip=half}}
\makeatother
\usepackage{xcolor}
\usepackage{longtable,booktabs,array}
\usepackage{calc} % for calculating minipage widths
\usepackage{etoolbox}
\makeatletter
\patchcmd\longtable{\par}{\if@noskipsec\mbox{}\fi\par}{}{}
\makeatother
\IfFileExists{footnotehyper.sty}{\usepackage{footnotehyper}}{\usepackage{footnote}}
\makesavenoteenv{longtable}
\providecommand{\tightlist}{%
  \setlength{\itemsep}{0pt}\setlength{\parskip}{0pt}}
\usepackage[margin=1in]{geometry}
\usepackage{newunicodechar}
\newunicodechar{−}{\ensuremath{-}}
\newunicodechar{×}{\ensuremath{\times}}
\newunicodechar{≤}{\ensuremath{\leq}}
\newunicodechar{≥}{\ensuremath{\geq}}
\newunicodechar{→}{\ensuremath{\rightarrow}}
\newunicodechar{≈}{\ensuremath{\approx}}
\newunicodechar{±}{\ensuremath{\pm}}
\newunicodechar{–}{--}
\newunicodechar{∩}{\ensuremath{\cap}}
\newunicodechar{⊆}{\ensuremath{\subseteq}}
\newunicodechar{Σ}{\ensuremath{\Sigma}}
\usepackage{etoolbox}
\AtBeginEnvironment{longtable}{\footnotesize\setlength{\tabcolsep}{4pt}}
\usepackage{caption}
\usepackage{microtype}
\usepackage{xurl}
\usepackage{needspace}

\ifLuaTeX
  \usepackage{selnolig}  % disable illegal ligatures
\fi
\IfFileExists{bookmark.sty}{\usepackage{bookmark}}{\usepackage{hyperref}}
\IfFileExists{xurl.sty}{\usepackage{xurl}}{} % add URL line breaks if available
\hypersetup{
  colorlinks=true,
  linkcolor={Maroon},
  filecolor={Maroon},
  citecolor={Blue},
  urlcolor={Blue},
  pdfcreator={LaTeX via pandoc}}

\author{}
\date{}

\begin{document}

\hypersetup{pdftitle={FERN: Cross-Benchmark Table Retrieval for Text-to-SQL with Frozen Retrievers and Learned Ranking}, pdfauthor={Arjun Rao, Nishant Gopinath}}
\title{FERN: Cross-Benchmark Table Retrieval for Text-to-SQL with Frozen Retrievers and Learned Ranking}
\author{Arjun Rao\\\texttt{\small arjun.rao01@sap.com} \and Nishant Gopinath\\\texttt{\small nishant.gopinath@sap.com}}
\date{November 2026}
\maketitle
\begin{abstract}
Text-to-SQL systems over large pooled catalogs must first retrieve the
tables a question needs. We compare against the published table recall
of RASL, which uses Claude 3.5 models to extract keywords and select
tables, on pooled BIRD, Spider, and Fiben catalogs. FERN
(Feature-Engineered Ranking with No generative LLM) instead fuses three
frozen retrievers into 100 candidates, describes each with up to 120
features, and reorders them with LightGBM rankers fitted once on BIRD's
training questions. On Spider, where it was not fitted, its candidate
pool alone reaches RASL's published recall at fifteen tables (98.0), and
the ranker raises the share of questions with every gold table in the
top five (complete@5) from 83.9\% to 95.6\%. At five tables, FERN's
recall is within one point of RASL's on Spider (96.7 vs 97.0) and on
BIRD with its evidence hints (96.6 vs 97.5), and 8.2 points below on
BIRD without them (89.3). On 121 real BEAVER questions, without BEAVER
fitting, its recall at fifteen (82.1) falls between two BEAVER-trained
retrievers (80.8, 84.5), but its complete@15 is below both (49.6 vs 54.5
and 62.0). On Fiben, it falls far short at fifteen tables (54.2 vs 69.2,
different question sets), although its recall at fifty is numerically
close to RASL's question-only retrieval result (80.7 vs 79.4). On BIRD
dev with evidence, non-retrieval features raise complete@5 by 7.7 points
over retrieval signals alone. On BIRD dev, median latency is 208 ms per
question on one A100, without hosted model calls.

\end{abstract}

\needspace{8\baselineskip}

\hypertarget{introduction}{%
\subsection{1
Introduction}\label{introduction}}

A text-to-SQL system
can write a faithful
query only if it has
the tables the query
needs. Over a pooled
catalog with
hundreds of tables
from many databases,
a retrieval stage
has to find those
few tables first: a
missing table makes
the query wrong, and
extra tables consume
the generator's
context. This step,
schema linking or
table retrieval,
increasingly uses
large language
models to extract
keywords, rewrite
questions, or read
candidate schemas
and choose (Pourreza
\& Rafiei, 2023;
Talaei et al., 2024;
Eben et al., 2025).
We use RASL's
primary
table-retrieval
results (Eben et
al., 2025) as a
published reference
on pooled BIRD,
Spider, and Fiben
catalogs. RASL
retrieves schema
entities with
keywords extracted
by Claude 3.5 Haiku
and selects tables
with Claude 3.5
Sonnet v2.

Generative linking
has costs beyond
accuracy. RASL uses
hosted models for
keyword extraction
and table selection;
its end-to-end
input-token estimate
is 1.85 US dollars
per 100 BIRD
queries, including 3
cents in the
category it labels
retrieval (Eben et
al., 2025). In
enterprise settings,
sending schema and
question text to an
external service may
not be acceptable.
Retrieval research
has long combined
many weak relevance
signals with a
learned ranker (Qin
\& Liu, 2013;
Burges, 2010), and
such a ranker needs
ingredients that can
run locally: lexical
search,
off-the-shelf
embedding models,
and schema
statistics.

This paper asks:
\textbf{how far can
a ranker trained
once on a public
benchmark take table
retrieval across
pooled catalogs,
using frozen
retrievers and no
generative model in
the retrieval path?}
We answer with FERN
(Feature-Engineered
Ranking with No
generative LLM).
Three frozen
retrievers propose
100 candidates; each
candidate is
described by
features computed
from the question,
the catalog, and the
retrievers' scores;
LightGBM rankers (Ke
et al., 2017) fitted
once on BIRD's
training questions
order them. We
compare with RASL's
published numbers on
the same pooled
catalogs, with no
database identifier
given, and apply the
BIRD-fitted rankers
unchanged to Spider,
Fiben, and three
pooled BEAVER
warehouses (Chen et
al., 2024b).

The results, in
brief. At a nominal
budget of five
tables, FERN's
recall is within 0.3
points of RASL's on
Spider and 0.9
points on BIRD with
evidence (BIRD's
per-question hints
that name relevant
columns and values);
when FERN returns
fifteen tables, it
is about one point
higher on both (on
question-only BIRD
it is 0.7 points
below RASL's BIRD
figure). Spider is
the cleanest of
these comparisons:
FERN was never
fitted on it, no
design decision used
it, and it involves
no evidence
ambiguity. Without
BIRD's evidence
hints, FERN's recall
at five falls by 7.3
points. FERN's own
candidate pool
already reaches
RASL's Spider recall
at fifteen (98.0);
the ranker's value
lies mainly at small
budgets and in
complete table sets.
On BEAVER, FERN's
recall at fifteen
falls between two
retrievers trained
on BEAVER, but fewer
questions get a
complete table set
(every gold table in
the list). On Fiben,
its recall is close
to RASL's
question-only
retrieval result at
fifty tables, but
substantially lower
than RASL's
selected-table
recall at fifteen
(Section 7).

\textbf{Contributions.}

\begin{enumerate}
\def\labelenumi{\arabic{enumi}.}
\tightlist
\item
  \textbf{FERN} is a
  table retriever
  that combines
  three frozen
  retrievers with
  gradient-boosted
  rankers over
  schema-specific
  features, in three
  variants for
  catalogs with or
  without evidence,
  cell values, and
  column types. It
  runs in about 0.2
  seconds per
  question on one
  GPU with no hosted
  model calls.
\item
  \textbf{A
  comparison with
  published
  LLM-based schema
  linking} (RASL) on
  its pooled BIRD,
  Spider, and Fiben
  catalogs. We made
  design choices on
  BIRD training data
  using rules stated
  in advance, with
  two documented
  exceptions; we
  fitted the rankers
  once; Spider and
  Fiben informed no
  design decisions.
\item
  \textbf{An
  analysis, in the
  frozen
  configuration, of
  which feature
  groups matter at
  which table
  budget.} On BIRD
  dev with evidence,
  rankers over
  retrieval signals
  alone or over the
  other features
  alone each reach
  about 98 recall at
  fifteen tables;
  adding the
  non-retrieval
  features to the
  retrieval signals
  raises complete@5
  by 7.7 points. In
  the exploratory
  complete@5
  ablations,
  removing
  database-support
  or join-graph
  features lowered
  the point estimate
  by 0.8 points
  each, with
  unadjusted
  intervals above
  zero; the other
  intervals touched
  or included zero.
\item
  \textbf{Transfer
  results} on a
  second benchmark,
  an opaque
  financial schema,
  and an enterprise
  benchmark with
  three warehouses,
  separating recall
  from complete-set
  recovery and
  showing, on these
  benchmarks, where
  a ranker trained
  on public data is
  enough and where
  it is not.
\end{enumerate}

\needspace{8\baselineskip}

\hypertarget{related-work}{%
\subsection{2
Related
work}\label{related-work}}

\textbf{Schema
linking with
language models.}
Earlier text-to-SQL
parsers linked
question spans to
schema elements by
name and value
matching inside the
model (Guo et al.,
2019; Wang, B. et
al., 2020). RESDSQL
ranks schema items
with a learned
cross-encoder before
decoding (Li, H. et
al., 2023), and
CodeS combines
schema-item
classification with
value retrieval (Li,
H. et al., 2024).
With large language
models, linking
became a prompting
step: DIN-SQL links
the schema with an
LLM before
generation (Pourreza
\& Rafiei, 2023) and
CHESS retrieves
entities and context
in an LLM-driven
pipeline (Talaei et
al., 2024). Maamari
et al.~(2024) argue
that strong
reasoning models
reduce the need for
explicit schema
linking when the
database is known
and small. These
methods mostly
assume a given
database. Our
setting pools many
databases without
identifying the
target, as in RASL
(Eben et al., 2025),
which indexes schema
entities with
descriptions,
retrieves with
LLM-extracted
keywords, calibrates
relevance on 200
training questions
per dataset where a
training split
exists (not Fiben),
and lets an LLM
choose the tables.
LinkAlign (Wang, Y.
et al., 2025) also
addresses linking
across many
databases.

\textbf{Retrieval
for large catalogs.}
CRUSH4SQL asks an
LLM to hallucinate a
minimal schema and
retrieves real
tables that match it
(Kothyari et al.,
2023). DBCopilot
routes questions to
databases and tables
with a generative
model trained on
synthetic questions
generated over the
target databases
(Wang, T. et al.,
2025); in RASL's
Table 2, it reports
94.6 recall at
fifteen on BIRD and
97.6 on Spider.
Dense table
retrieval embeds
questions and tables
in a shared space
(Herzig et al.,
2021). Chen, Zhang
\& Roth (2024a) show
that join-aware
retrieval matters
when questions need
several tables.
CORE-T (Soliman et
al., 2026) targets
coherent multi-table
retrieval over
pooled catalogs
without database
identifiers and also
evaluates on BEAVER;
it reports
table-selection F1
rather than recall@k
on RASL's catalogs,
so we do not compare
with it numerically.
CSR-RAG (Singh et
al., 2026) combines
similarity to past
questions on the
target catalog, a
schema knowledge
graph, and
hypergraph ranking,
without an LLM in
retrieval; FERN uses
no target-catalog
history and is
fitted once on a
public benchmark.
NIFT-SQL (Rao \&
Gopinath, 2026)
fine-tunes a
necessity scorer and
a retriever on a
customer's own
questions and fuses
them; it asks what
target-catalog
supervision buys.
FERN asks the
complementary
question: what a
ranker fitted once
on public data
delivers with no
target-domain
fitting. Both use
database-level
support features in
a learned fusion;
FERN differs in
using only frozen
retrievers, a larger
schema-specific
feature set,
cross-benchmark
transfer, and a
comparison with
LLM-based linking.

\textbf{Learning to
rank and hybrid
retrieval.} Learning
to rank over
hand-built relevance
features is classic
information
retrieval (Qin \&
Liu, 2013), with
gradient-boosted
trees such as
LambdaMART among the
strongest rankers on
such features
(Burges, 2010; Qin,
Z. et al., 2021).
For tables, Zhang \&
Balog (2018) already
rank web tables with
supervised learning
over lexical and
semantic features;
FERN differs in its
target (the set of
tables a SQL query
needs), its pooled
multi-database
catalogs, and its
evaluation of
transfer across
benchmarks. Hybrid
lexical and dense
retrieval is a
standard first
stage, commonly
fused by reciprocal
rank fusion (Cormack
et al., 2009; Bruch
et al., 2023). We
claim no new ranking
algorithm. The
contribution is
empirical: how far
this recipe reaches
against LLM-based
linking, at which
budgets, and where
it stops.

\needspace{8\baselineskip}

\hypertarget{problem-setting}{%
\subsection{3
Problem
setting}\label{problem-setting}}

\textbf{Task.} Given
a natural-language
question and a
pooled catalog of
tables from many
databases, return a
ranked list of
tables such that the
tables the gold SQL
query uses appear
near the top. The
target database is
not given.

\textbf{Catalogs.}
Table 1 lists the
catalogs. For BIRD
(Li, J. et al.,
2023) and Spider (Yu
et al., 2018), we
pool every database
in the training and
development splits
into one catalog, as
in RASL; Fiben (Sen
et al., 2020) is a
single financial
database. Our table
and column counts
for BIRD, Spider,
and Fiben match
RASL's Table 1.
BEAVER's current
public release pools
three warehouses
(dw, nova, neutron);
its 121 real
questions come from
query logs of the dw
warehouse, and we
score a random
sample (seed 7) of
1,000 of its 7,857
synthetic questions.

\needspace{15\baselineskip}

\emph{Table 1.
Catalogs and test
questions.}

\begin{longtable}[]{@{}
  >{\raggedright\arraybackslash}p{(\columnwidth - 10\tabcolsep) * \real{0.1967}}
  >{\raggedright\arraybackslash}p{(\columnwidth - 10\tabcolsep) * \real{0.1803}}
  >{\raggedright\arraybackslash}p{(\columnwidth - 10\tabcolsep) * \real{0.1311}}
  >{\raggedright\arraybackslash}p{(\columnwidth - 10\tabcolsep) * \real{0.1475}}
  >{\raggedright\arraybackslash}p{(\columnwidth - 10\tabcolsep) * \real{0.1803}}
  >{\raggedright\arraybackslash}p{(\columnwidth - 10\tabcolsep) * \real{0.1639}}@{}}
\toprule\noalign{}
\begin{minipage}[b]{\linewidth}\raggedright
Dataset
\end{minipage} &
\begin{minipage}[b]{\linewidth}\raggedright
Databases
\end{minipage} &
\begin{minipage}[b]{\linewidth}\raggedright
Tables
\end{minipage} &
\begin{minipage}[b]{\linewidth}\raggedright
Columns
\end{minipage} &
\begin{minipage}[b]{\linewidth}\raggedright
Test questions
\end{minipage} &
\begin{minipage}[b]{\linewidth}\raggedright
Gold tables per
question
\end{minipage} \\
\midrule\noalign{}
\endhead
\bottomrule\noalign{}
\endlastfoot
BIRD dev & 80 & 597
& 4,337 & 1,534 &
1.94 \\
Spider dev & 166 &
876 & 4,503 & 1,034
& 1.51 \\
Fiben & 1 & 152 &
374 & 298 (RASL:
279) & 3.82 \\
BEAVER, three
warehouses & 3 & 381
& & 121 real; 1,000
synthetic & 3.93;
3.52 \\
\end{longtable}

\textbf{Gold
tables.} The gold
tables of a question
are the catalog
tables, within the
question's database,
referenced in its
gold SQL, extracted
with the sqlglot
parser (sqlglot,
2024); names
introduced by common
table expressions
are excluded.
Fiben's gold SQL is
parsed as
PostgreSQL, the
others as SQLite. On
Fiben, 298 of 300
questions parse.
RASL reports 279
Fiben questions
without listing
which, so the Fiben
comparison uses
different question
sets (Section 6.1
bounds the effect).
BEAVER provides gold
table lists.
Extraction matters
at this margin: a
table-name regular
expression instead
of the parser
changes the gold set
for 25 of 1,534 BIRD
dev questions
(1.6\%), all because
a table named
\texttt{order}
matches the keyword
\texttt{ORDER\ BY}.

\textbf{Metrics.}
For a question with
gold set \(G\) and
top-k list \(T_k\),
recall@k is
\(|G \cap T_k| / |G|\),
macro-averaged over
questions as in
RASL. We also report
complete@k, the
share of questions
with
\(G \subseteq T_k\),
because a query that
misses any required
table is wrong. A
complete table set
is necessary for a
correct query but
does not guarantee
one. RASL's N is a
maximum budget: its
LLM selects only the
tables it judges
relevant, and on
BIRD and Spider it
never predicts more
than 15 (Eben et
al., 2025). FERN
returns exactly
fifteen tables at k
= 15, whereas RASL
may return fewer.
Comparisons at five
and fifteen
therefore use the
same nominal caps
but may yield
different realized
output sizes.
Intervals are 95\%
percentile bootstrap
intervals over
questions (2,000
resamples, seed 0);
Section 6.1 adds
intervals that
resample whole
databases. They
describe uncertainty
only in our
estimates. RASL's
per-question results
are not available,
so differences from
RASL are point
differences, not
tests.

\needspace{22\baselineskip}

\emph{Table 2.
Inputs available to
each system.
FERN-ev, FERN-q and
FERN-schema are
defined in Section
4.}

\begin{longtable}[]{@{}
  >{\raggedright\arraybackslash}p{(\columnwidth - 8\tabcolsep) * \real{0.1833}}
  >{\raggedright\arraybackslash}p{(\columnwidth - 8\tabcolsep) * \real{0.2000}}
  >{\raggedright\arraybackslash}p{(\columnwidth - 8\tabcolsep) * \real{0.3167}}
  >{\raggedright\arraybackslash}p{(\columnwidth - 8\tabcolsep) * \real{0.1167}}
  >{\raggedright\arraybackslash}p{(\columnwidth - 8\tabcolsep) * \real{0.1833}}@{}}
\toprule\noalign{}
\begin{minipage}[b]{\linewidth}\raggedright
\end{minipage} &
\begin{minipage}[b]{\linewidth}\raggedright
RASL full
\end{minipage} &
\begin{minipage}[b]{\linewidth}\raggedright
FERN-ev
\end{minipage} &
\begin{minipage}[b]{\linewidth}\raggedright
FERN-q
\end{minipage} &
\begin{minipage}[b]{\linewidth}\raggedright
FERN-schema
\end{minipage} \\
\midrule\noalign{}
\endhead
\bottomrule\noalign{}
\endlastfoot
Supervision & 200
calibration
questions per
dataset (none on
Fiben) & 9,394
labeled BIRD train
questions, fitted
once & same &
same \\
BIRD evidence & not
stated; its examples
include
evidence-like
clauses & yes & no &
no \\
Cell values & not
indexed & yes & yes
& no \\
Column and value
descriptions (BIRD)
& yes & no & no &
no \\
Generative model
calls per question &
2 (keywords, table
selection) & 0 & 0 &
0 \\
Embedding models &
Cohere Embed v3 &
Qwen3-Embedding-0.6B,
BGE-large & same &
same \\
Output size & at
most N, chosen by
the LLM & exactly k
& exactly k &
exactly k \\
\end{longtable}

\needspace{8\baselineskip}

\hypertarget{method}{%
\subsection{4
Method}\label{method}}

FERN has four
stages.

\begin{enumerate}
\def\labelenumi{\arabic{enumi}.}
\tightlist
\item
  \textbf{Frozen
  retrieval.} Each
  table is
  represented by a
  full card,
  \texttt{table\ (database\ db):\ col1,\ col2\ (natural\ name),\ ...},
  where underscores
  in the database
  name become spaces
  and a column's
  natural-language
  name (BIRD's and
  Spider's
  \texttt{column\_\allowbreak{}names})
  is added in
  parentheses when
  it differs from
  the stored name
  after removing
  spaces,
  underscores, and
  case; and by a
  name card,
  \texttt{table\ words\ (database\ db)},
  with camelCase and
  underscores split.
  Cards contain no
  types, keys, or
  values and are not
  truncated;
  BGE-large reads at
  most 512 tokens,
  so for very wide
  tables its signal
  sees only the
  first part of the
  card. Five signals
  score every table:
  BM25 (Robertson \&
  Zaragoza, 2009) on
  full cards and on
  name cards
  (rank\_\allowbreak{}bm25
  BM25Okapi, k1 =
  1.5, b = 0.75;
  tokens are
  lower-cased
  alphanumeric runs
  after splitting
  camelCase and
  underscores, with
  no stemming or
  stop words);
  Qwen3-Embedding-0.6B
  (Zhang, Y. et al.,
  2025) on full and
  name cards, with
  its built-in query
  instruction for
  questions; and
  BGE-large-en-v1.5
  (Xiao et al.,
  2024) on full
  cards, with its
  retrieval prefix
  for questions.
  Embeddings are
  L2-normalized and
  compared by dot
  product, in
  bfloat16. The
  question text is
  the question
  followed by BIRD's
  evidence in
  FERN-ev and the
  question alone
  otherwise.
\item
  \textbf{Candidate
  pool.} The pool
  score of a table
  is
  \(\sum_s 1/(60 + \mathrm{rank}_s)\),
  summed over BM25,
  Qwen3, and BGE on
  full cards, with
  zero-based ranks
  over the whole
  catalog (Cormack
  et al., 2009). The
  100 tables with
  the highest pool
  score are the
  candidates
  (average recall at
  100 is 99.4 on
  BIRD dev). Ties
  are broken by
  array order.
\item
  \textbf{Features.}
  Table 3 describes
  each candidate's
  features. We
  compute ranks,
  z-scores,
  percentiles, and
  gaps over the 100
  candidates for
  each question.
\item
  \textbf{Ranking.}
  A LightGBM ranker
  with the
  lambdarank
  objective, trained
  with each
  question's
  candidates as one
  group (pairwise
  gradients) and
  applied pointwise,
  scores the
  candidates, which
  are then sorted by
  that score.
\end{enumerate}

\needspace{31\baselineskip}

\emph{Table 3.
Features. The six
signals in the score
and agree families
are the pool score,
BM25 (full and name
cards), Qwen3 (full
and name cards), and
BGE. FERN-ev uses
all 120.}

\begin{longtable}[]{@{}
  >{\raggedright\arraybackslash}p{(\columnwidth - 4\tabcolsep) * \real{0.0833}}
  >{\raggedright\arraybackslash}p{(\columnwidth - 4\tabcolsep) * \real{0.0500}}
  >{\raggedright\arraybackslash}p{(\columnwidth - 4\tabcolsep) * \real{0.8667}}@{}}
\toprule\noalign{}
\begin{minipage}[b]{\linewidth}\raggedright
Family
\end{minipage} &
\begin{minipage}[b]{\linewidth}\raggedright
n
\end{minipage} &
\begin{minipage}[b]{\linewidth}\raggedright
Content
\end{minipage} \\
\midrule\noalign{}
\endhead
\bottomrule\noalign{}
\endlastfoot
score & 30 & each
signal's raw score,
rank, z-score, gap
to the best
candidate and
percentile \\
agree & 9 & minimum,
maximum, mean and
standard deviation
of the six ranks;
how many signals
place the table in
their top 5 and top
15; rank differences
Qwen3 minus BM25,
Qwen3 minus BGE,
name card minus full
card \\
db & 14 & for the
pool score and for
Qwen3: the
database's best
score, mean score
and probability
share (softmax of
the candidates'
z-scores, summed by
database), the
database's rank by
best score, the
table's rank within
its database, its
gap to the
database's best;
plus the number of
candidates from the
database and its
table count \\
lex & 8 & overlap of
question tokens with
table-name tokens
and with all name
tokens of the table,
IDF-weighted
overlap, overlap
count, best fuzzy
match, fuzzy matches
above 0.8, natural
column names found
in the question,
table name found in
the question \\
cov & 3 & share of
question tokens
covered by the
table's tokens;
tokens it adds
beyond the top
candidate and beyond
the top three \\
ev & 3 & evidence
contains the table
name, how many
column names, how
many natural column
names (zero without
evidence) \\
val & 6 & question
literals found among
the table's cell
values; share of
literals matched;
columns with a
match; literals
matched only by this
table within its
database; question
numbers inside a
numeric column's
minimum-maximum
range; share in
range \\
type & 7 & numeric
columns, share
numeric, date-like
columns; comparison,
aggregation, time
and percentage
questions crossed
with numeric or
date-like columns \\
meta & 6 & columns,
primary-key columns,
foreign-key columns,
log(1 + rows),
junction table (at
least two
foreign-key columns
and at most four
columns), name
length \\
graph & 7 &
foreign-key degree,
PageRank, join
distance to the top
candidate, neighbors
among the top five,
bridge (neighbor of
two or more of the
top five), shared
non-generic column
names with the top
candidate and with
the top five \\
q & 11 & question
words, content
tokens, literals,
numbers, presence of
evidence, entropy of
the softmax of
pool-score z-scores,
gap between the top
two pool scores,
aggregation,
comparison,
percentage and time
flags \\
emb & 16 &
element-wise product
of the Qwen3
question and
full-card
embeddings,
projected to 16
dimensions \\
\end{longtable}

\textbf{Feature
details.} Tokens for
lex, cov, and IDF
are lower-cased
alphanumeric runs
after camelCase and
underscore
splitting, with a
fixed list of about
60 function words
removed and a final
s stripped from
words longer than
three characters;
IDF is log(N / (1 +
df)) over the
catalog's tables.
Fuzzy matching
compares each
question token of at
least four
characters with
every catalog token
by normalized Indel
similarity
(rapidfuzz).
Literals are quoted
strings of 2 to 60
characters,
capitalized spans of
the question with
leading function
words removed, and
numbers. Cell values
are indexed once per
catalog from the
databases' SQLite
files: up to 20,000
distinct non-null
values per text
column, lower-cased,
2 to 60 characters,
matched exactly; and
the minimum and
maximum of numeric
columns. A column is
date-like if its
type contains date
or time, or its name
contains date, year,
or time; fixed
regular expressions
detect question
types. PageRank
(Page et al., 1999;
damping 0.85, 30
iterations) and join
distances (up to
five hops, nine
otherwise) use
declared foreign
keys, undirected,
within each
database. The top
candidate has the
highest pool score.
The projection for
emb is a PCA with 16
components fitted on
every ninth
candidate row of
BIRD train (with
evidence) and reused
for all variants. We
will release the
complete feature
code with the
materials listed
under Data and code
availability. The
pipeline is
deterministic:
rerunning it from
scratch reproduced
every reported
retrieval number
exactly.

\textbf{Variants and
fitting.} Three
rankers are fitted
once on 9,394 BIRD
training questions
whose gold SQL
parses and names a
catalog table (RASL
reports 9,427
training questions).
FERN-ev uses the
question with
evidence and all 120
features. FERN-q
uses the question
alone and drops the
evidence features
and the evidence
flag (116).
FERN-schema also
drops cell values,
column types, and
row counts (102) for
catalogs whose
values or types are
not available to us.
Labels are 1 for
gold tables among
the candidates and 0
otherwise. The
authors set
hyperparameters once
before the
development runs,
using common
LightGBM settings,
and never tuned
them: 600 trees,
learning rate 0.05,
63 leaves, at least
50 rows per leaf,
row subsampling 0.8
every iteration,
column subsampling
0.7, L2
regularization 1.0,
seed 0, other
parameters at
LightGBM defaults.

\textbf{Which
variant runs where.}
BIRD with evidence
uses FERN-ev; BIRD
without evidence and
Spider use FERN-q
(Spider has values
but no evidence).
Fiben and BEAVER use
FERN-schema: our
copies have no cell
values, BEAVER
provides no column
types, and Fiben's
declared foreign
keys (159) and
BEAVER's declared
keys, where they
exist, feed the
graph features.

\needspace{8\baselineskip}

\hypertarget{protocol}{%
\subsection{5
Protocol}\label{protocol}}

\textbf{Rules stated
in advance.} The
candidate pool and
the set of signals
were chosen by rules
fixed before running
them and applied
with five-fold
cross-validation
(folds by database)
on a 3,000-question
subset of BIRD
train: the pool with
the highest
complete@100, with
ties within 0.2
points going to
fewer retrievers;
and a signal kept
only if removing it
costs at least 0.3
points of complete@5
or complete@15, with
a joint-removal
check. The rule
removed a
best-column dense
signal (Appendix A).

\textbf{Two
documented
exceptions.} (1) The
pool rule selects
Qwen3 + BGE
(complete@100 98.4)
over BM25 + Qwen3 +
BGE (98.3). We
overrode it to
include lexical
retrieval, arguing
that exact matching
complements dense
retrieval for codes
and abbreviations.
The override was
made after an
earlier BEAVER
transfer check, in a
different
configuration, had
been seen, but
before Spider and
Fiben were scored;
it should be read as
informed by that
check. (2) A
fine-tuned decision
model was evaluated
on BIRD dev during
development and left
out of FERN because
it added little
(Appendix B). That
decision used BIRD
dev.

\textbf{What each
test set has
informed.} Spider
and Fiben informed
no decision. Fiben
was evaluated with
FERN-schema. Spider
was evaluated with
FERN-q and, after
the method was
frozen, with
FERN-schema for the
analysis without
cell values in
Section 6.1; neither
Spider evaluation
informed a design
decision. That
analysis also uses a
fourth ranker,
FERN-ev refitted on
BIRD train without
cell values, column
types, or row
counts. BIRD dev
informed the second
exception and the
development studies
of Appendix B.
BEAVER informed the
first exception.
BIRD dev and BEAVER
results are
therefore
development-exposed;
Spider is the
cleanest comparison
with RASL. We
exclude evaluation
questions and labels
from fitting; each
question is
available at
inference, and
catalogs are visible
as the task
requires: BM25
statistics, IDF, and
the value index are
built over the
catalog being
searched.

\textbf{Comparison
with RASL.} We take
RASL's retrieval
results from its
published Tables 2
and 5 and its cost
figures from its
cost analysis (Eben
et al., 2025); we
did not rerun it. We
use the same named
development splits
and reported catalog
counts for BIRD and
Spider; exact
question-ID and
gold-label parity
with RASL is not
verified. Fiben uses
a different question
set. Table 2 lists
the input
differences.

\textbf{Leakage, by
provenance.} Table 4
lists every input
and the stages that
see it. The feature
code receives the
question, the
evidence (FERN-ev
only), the catalog,
the cell values, and
the frozen scores;
candidate identities
come from the whole
catalog, and the
gold SQL, gold
tables, and gold
database are never
passed to it. Labels
are attached after
features are
computed. PCA and
the rankers are
fitted only on BIRD
train. As a coarse
sanity check, not a
proof, no single
feature separates
gold from non-gold
candidates on BIRD
dev with an AUC
above 0.98; the
strongest, the pool
score's gap to the
best candidate,
reaches 0.978, as
expected of the main
retrieval signal.

\needspace{19\baselineskip}

\emph{Table 4.
Inputs and the
stages that use
them.}

\begin{longtable}[]{@{}
  >{\raggedright\arraybackslash}p{(\columnwidth - 4\tabcolsep) * \real{0.2034}}
  >{\raggedright\arraybackslash}p{(\columnwidth - 4\tabcolsep) * \real{0.2712}}
  >{\raggedright\arraybackslash}p{(\columnwidth - 4\tabcolsep) * \real{0.5254}}@{}}
\toprule\noalign{}
\begin{minipage}[b]{\linewidth}\raggedright
Input
\end{minipage} &
\begin{minipage}[b]{\linewidth}\raggedright
Source
\end{minipage} &
\begin{minipage}[b]{\linewidth}\raggedright
Used by
\end{minipage} \\
\midrule\noalign{}
\endhead
\bottomrule\noalign{}
\endlastfoot
Question text & the
benchmark question &
retrievers,
features \\
Evidence text &
BIRD's evidence
field & FERN-ev
only: appended to
the question for all
retrievers and for
every feature
computed from
question text (lex,
cov, val literals,
type and q flags,
emb), plus the ev
features and the
evidence flag \\
Schema: names,
types, keys & the
catalog & retrievers
(names only),
features (names,
types and keys) \\
Cell values, row
counts & the
catalog's SQLite
files, never the
gold SQL & val and
meta features,
FERN-ev and
FERN-q \\
Retriever scores &
frozen models &
score, agree and db
features \\
Gold tables & parsed
from gold SQL for
BIRD, Spider and
Fiben; supplied by
BEAVER for BEAVER &
fitting the rankers
(BIRD train) and
evaluation only \\
\end{longtable}

\needspace{8\baselineskip}

\hypertarget{results}{%
\subsection{6
Results}\label{results}}

\needspace{8\baselineskip}

\hypertarget{comparison-with-published-llm-based-linking}{%
\subsubsection{6.1
Comparison with
published LLM-based
linking}\label{comparison-with-published-llm-based-linking}}

\needspace{27\baselineskip}

\emph{Table 5. Table
recall on the pooled
catalogs. RASL full
uses Claude 3.5
Haiku keywords and
Claude 3.5 Sonnet v2
table selection and
returns at most N
tables; RASL
retriever ranks by
retrieval with the
Haiku keywords,
without LLM table
selection. The BM25,
DBCopilot, RASL
retriever, and RASL
full rows are from
RASL's Table 2.
Differences are FERN
minus RASL full, in
points. FERN
intervals: 95\%
bootstrap over
questions.}

\begin{longtable}[]{@{}
  >{\raggedright\arraybackslash}p{(\columnwidth - 12\tabcolsep) * \real{0.1864}}
  >{\raggedright\arraybackslash}p{(\columnwidth - 12\tabcolsep) * \real{0.1356}}
  >{\raggedright\arraybackslash}p{(\columnwidth - 12\tabcolsep) * \real{0.1356}}
  >{\raggedright\arraybackslash}p{(\columnwidth - 12\tabcolsep) * \real{0.1356}}
  >{\raggedright\arraybackslash}p{(\columnwidth - 12\tabcolsep) * \real{0.1356}}
  >{\raggedright\arraybackslash}p{(\columnwidth - 12\tabcolsep) * \real{0.1356}}
  >{\raggedright\arraybackslash}p{(\columnwidth - 12\tabcolsep) * \real{0.1356}}@{}}
\toprule\noalign{}
\begin{minipage}[b]{\linewidth}\raggedright
\end{minipage} &
\begin{minipage}[b]{\linewidth}\raggedright
BIRD R@5
\end{minipage} &
\begin{minipage}[b]{\linewidth}\raggedright
BIRD R@15
\end{minipage} &
\begin{minipage}[b]{\linewidth}\raggedright
Spider R@5
\end{minipage} &
\begin{minipage}[b]{\linewidth}\raggedright
Spider R@15
\end{minipage} &
\begin{minipage}[b]{\linewidth}\raggedright
Fiben R@5
\end{minipage} &
\begin{minipage}[b]{\linewidth}\raggedright
Fiben R@15
\end{minipage} \\
\midrule\noalign{}
\endhead
\bottomrule\noalign{}
\endlastfoot
BM25 (RASL's Table
2) & 68.3 & 82.8 &
86.5 & 93.9 & 33.3 &
38.6 \\
DBCopilot & 85.8 &
94.6 & 91.6 & 97.6 &
41.1 & 56.9 \\
RASL retriever &
70.5 & 92.6 & 72.1 &
94.1 & 34.6 &
64.5 \\
RASL full & 97.5 &
97.8 & 97.0 & 98.0 &
69.1 & 69.2 \\
FERN pool alone &
86.8 & 96.2 & 89.6 &
98.0 & 24.8 &
39.3 \\
FERN & 96.6 {[}95.9,
97.4{]} & 98.9
{[}98.4, 99.2{]} &
96.7 {[}95.6,
97.6{]} & 99.1
{[}98.6, 99.6{]} &
36.7 {[}33.7,
39.5{]} & 54.2
{[}50.8, 57.4{]} \\
FERN minus RASL full
& -0.9 & +1.1 & -0.3
& +1.1 & -32.4 &
-15.0 \\
FERN, BIRD question
only & 89.3 {[}88.0,
90.6{]} & 97.1
{[}96.4, 97.7{]} & &
& & \\
\end{longtable}

FERN variants:
FERN-ev on BIRD,
FERN-q on
question-only BIRD
and on Spider,
FERN-schema on
Fiben. Fiben: FERN
298 questions, RASL
279.

At a nominal budget
of five tables, FERN
is 0.3 points below
RASL's published
recall on Spider and
0.9 points below on
BIRD with evidence.
When FERN returns
fifteen tables, it
is 1.1 points higher
on both; RASL
returns at most
fifteen, so this
compares a full list
of fifteen with a
possibly shorter
one. Whether RASL
used BIRD's evidence
is not stated: its
error-analysis
examples (Figures 3
and 5 and Appendix E
of Eben et al.,
2025) show questions
with evidence-like
clauses appended,
which supports
FERN-ev as the
closer comparison.
Without evidence,
FERN's recall at
five falls to 89.3,
8.2 points below
RASL, and its recall
at fifteen to 97.1,
0.7 points below.
For deployments
without per-question
evidence hints, the
question-only row is
more relevant.

FERN's candidate
pool alone already
reaches RASL's
Spider recall at
fifteen (98.0) and
is above every
baseline without LLM
table selection in
RASL's comparison at
fifteen tables, on
Spider and on BIRD
with evidence (the
question-only BIRD
pool, 94.2, is below
DBCopilot's 94.6).
Much of the gap
between earlier
non-LLM baselines
and RASL is thus
closed by the frozen
candidate pool; the
ranker earns its
place at small
budgets and on
complete sets (Table
6). The BM25 rows of
Tables 6 and 8 rank
FERN's 100
candidates by their
BM25 score; they
show what BM25
contributes as a
signal, not BM25 as
a standalone
retriever, and are
not comparable with
RASL's BM25 row.
Over the whole
catalog, our
word-level BM25 is
still weaker than
the BM25 row of
RASL's Table 2,
which RASL adopts
from DBCopilot's
reported results
with BM25 parameters
optimized on
training data (Eben
et al., 2025,
Appendix F); we did
not tune ours.

On Fiben, we
evaluate the two
systems on different
question sets. If
RASL's 279 questions
are a subset of our
298, FERN's recall
on them is bounded
by the two extremes
for the 19 others:
51.1 to 57.9 at
fifteen and 32.4 to
39.2 at five, far
below RASL's 69.2
and 69.1 in either
case.

Questions are
clustered in
databases, so we
also resample whole
databases (11 on
BIRD dev, 20 on
Spider dev). The
intervals widen to
{[}94.3, 98.5{]} for
BIRD recall at five
and {[}97.7, 99.7{]}
at fifteen, and to
{[}94.2, 98.3{]} and
{[}98.3, 99.7{]} on
Spider. These are
approximate given so
few clusters, and
they describe the
sensitivity of
FERN's own
estimates; they do
not test the
difference from
RASL, which we
report only as
signed point
differences.

\needspace{31\baselineskip}

\emph{Table 6.
Recall and
complete@k against
the candidate pool
on the same
catalogs. ``BM25
rerank of pool''
ranks the 100
candidates by BM25
score. C@k is
complete@k.}

\begin{longtable}[]{@{}
  >{\raggedright\arraybackslash}p{(\columnwidth - 14\tabcolsep) * \real{0.1695}}
  >{\raggedright\arraybackslash}p{(\columnwidth - 14\tabcolsep) * \real{0.2203}}
  >{\raggedright\arraybackslash}p{(\columnwidth - 14\tabcolsep) * \real{0.1017}}
  >{\raggedright\arraybackslash}p{(\columnwidth - 14\tabcolsep) * \real{0.1017}}
  >{\raggedright\arraybackslash}p{(\columnwidth - 14\tabcolsep) * \real{0.1017}}
  >{\raggedright\arraybackslash}p{(\columnwidth - 14\tabcolsep) * \real{0.1017}}
  >{\raggedright\arraybackslash}p{(\columnwidth - 14\tabcolsep) * \real{0.1017}}
  >{\raggedright\arraybackslash}p{(\columnwidth - 14\tabcolsep) * \real{0.1017}}@{}}
\toprule\noalign{}
\begin{minipage}[b]{\linewidth}\raggedright
Dataset
\end{minipage} &
\begin{minipage}[b]{\linewidth}\raggedright
Method
\end{minipage} &
\begin{minipage}[b]{\linewidth}\raggedright
R@5
\end{minipage} &
\begin{minipage}[b]{\linewidth}\raggedright
R@10
\end{minipage} &
\begin{minipage}[b]{\linewidth}\raggedright
R@15
\end{minipage} &
\begin{minipage}[b]{\linewidth}\raggedright
C@5
\end{minipage} &
\begin{minipage}[b]{\linewidth}\raggedright
C@10
\end{minipage} &
\begin{minipage}[b]{\linewidth}\raggedright
C@15
\end{minipage} \\
\midrule\noalign{}
\endhead
\bottomrule\noalign{}
\endlastfoot
BIRD, evidence &
BM25 rerank of pool
& 50.1 & 66.0 & 73.6
& 31.4 & 47.0 &
56.1 \\
& pool & 86.8 & 93.6
& 96.2 & 75.0 & 87.4
& 91.9 \\
& FERN-ev & 96.6 &
98.5 & 98.9 & 93.9 &
96.8 & 97.7 \\
BIRD, question only
& BM25 rerank of
pool & 30.0 & 48.1 &
58.5 & 16.3 & 30.4 &
40.1 \\
& pool & 81.5 & 90.4
& 94.2 & 67.4 & 81.6
& 88.3 \\
& FERN-q & 89.3 &
95.1 & 97.1 & 83.3 &
91.9 & 95.0 \\
Spider & BM25 rerank
of pool & 65.6 &
74.2 & 78.3 & 56.3 &
66.4 & 71.7 \\
& pool & 89.6 & 96.6
& 98.0 & 83.9 & 94.2
& 96.7 \\
& FERN-q & 96.7 &
98.4 & 99.1 & 95.6 &
98.2 & 98.9 \\
Fiben & BM25 rerank
of pool & 25.4 &
34.6 & 39.4 & 1.7 &
3.4 & 4.0 \\
& pool & 24.8 & 34.2
& 39.3 & 0.7 & 2.7 &
4.0 \\
& FERN-schema & 36.7
& 44.4 & 54.2 & 3.7
& 7.4 & 15.1 \\
\end{longtable}

Reranking the same
pool raises
complete@5 by 18.9
points on BIRD with
evidence (75.0 to
93.9) and by 11.7 on
Spider (83.9 to
95.6), and
complete@15 by 5.8
and 2.2.

\needspace{28\baselineskip}

\emph{Table 7.
Recall@k and
complete@k for k = 1
to 8, 10, 15 and 20
(R / C).}

\begin{longtable}[]{@{}
  >{\raggedright\arraybackslash}p{(\columnwidth - 8\tabcolsep) * \real{0.1000}}
  >{\raggedright\arraybackslash}p{(\columnwidth - 8\tabcolsep) * \real{0.2500}}
  >{\raggedright\arraybackslash}p{(\columnwidth - 8\tabcolsep) * \real{0.2167}}
  >{\raggedright\arraybackslash}p{(\columnwidth - 8\tabcolsep) * \real{0.2500}}
  >{\raggedright\arraybackslash}p{(\columnwidth - 8\tabcolsep) * \real{0.1833}}@{}}
\toprule\noalign{}
\begin{minipage}[b]{\linewidth}\raggedright
k
\end{minipage} &
\begin{minipage}[b]{\linewidth}\raggedright
BIRD, FERN-ev
\end{minipage} &
\begin{minipage}[b]{\linewidth}\raggedright
BIRD, FERN-q
\end{minipage} &
\begin{minipage}[b]{\linewidth}\raggedright
Spider, FERN-q
\end{minipage} &
\begin{minipage}[b]{\linewidth}\raggedright
BIRD pool alone (R)
\end{minipage} \\
\midrule\noalign{}
\endhead
\bottomrule\noalign{}
\endlastfoot
1 & 54.7 / 21.6 &
45.8 / 18.1 & 64.6 /
44.8 & 47.6 \\
2 & 84.4 / 69.7 &
70.7 / 51.6 & 86.2 /
77.4 & 68.9 \\
3 & 92.5 / 85.5 &
81.4 / 69.8 & 93.2 /
89.7 & 78.1 \\
4 & 95.4 / 91.3 &
86.3 / 78.3 & 95.6 /
93.9 & 84.0 \\
5 & 96.6 / 93.9 &
89.3 / 83.3 & 96.7 /
95.6 & 86.8 \\
6 & 97.4 / 95.4 &
91.0 / 85.7 & 96.9 /
95.9 & 89.1 \\
7 & 97.8 / 95.9 &
92.6 / 88.2 & 97.1 /
96.2 & 90.8 \\
8 & 98.2 / 96.4 &
93.6 / 89.6 & 97.5 /
97.0 & 92.0 \\
10 & 98.5 / 96.8 &
95.1 / 91.9 & 98.4 /
98.2 & 93.6 \\
15 & 98.9 / 97.7 &
97.1 / 95.0 & 99.1 /
98.9 & 96.2 \\
20 & 99.1 / 98.4 &
97.8 / 96.1 & 99.3 /
99.1 & 97.0 \\
\end{longtable}

FERN first reaches
or exceeds RASL's
published recall at
five at k = 7, on
Spider and on BIRD
with evidence. On
BIRD without
evidence, it is
below that reference
at k = 15 and above
it at k = 20.
Because realized
list sizes and
inputs differ, this
is an operational
comparison, not a
measure of what LLM
selection adds.

\textbf{Without cell
values, column
types, or row
counts.} RASL does
not index database
values, while
FERN-ev and FERN-q
do. Removing values,
column types, and
row counts together,
with separately
fitted rankers,
changes recall at
five and fifteen by
0.3 and 0.2 points
on BIRD with
evidence (FERN-ev
refitted, 106
features: 96.3 and
98.7) and by 0.8 and
0.1 on Spider (the
already-fitted
FERN-schema: 95.9
and 99.0). FERN's
results are
therefore robust to
removing these
inputs together;
this does not
establish parity
with RASL's schema
representation or
evidence use.
Without them, FERN
is 1.2 points below
RASL's published
recall at five
tables on BIRD and
1.1 on Spider.

\needspace{8\baselineskip}

\hypertarget{transfer-to-an-enterprise-benchmark}{%
\subsubsection{6.2
Transfer to an
enterprise
benchmark}\label{transfer-to-an-enterprise-benchmark}}

BEAVER (Chen et al.,
2024b) is built from
real enterprise data
warehouses. We apply
FERN-schema as
fitted on BIRD; no
BEAVER questions are
used for fitting.
The comparison rows
come from NIFT-SQL's
confirmation run on
the same release and
the same 121 real
questions from the
dw warehouse (Rao \&
Gopinath, 2026), and
both are trained on
BEAVER's synthetic
questions. NIFT-SQL
fuses a necessity
scorer with hybrid
search over a
retriever fine-tuned
on BEAVER; the
second row is the
retriever fine-tuned
on the same training
and validation
questions, the
label-matched
baseline that
NIFT-SQL reports. In
that run, the full
system recovered
fewer complete sets
on the real
questions than this
retriever alone
(54.5 vs 62.0). From
the NIFT-SQL study,
we reuse only its
prepared BEAVER data
files, its
declared-key list,
and these two
comparison rows; the
candidate pool,
features, rankers,
and all other
experiments are new.
Because we do not
have NIFT-SQL's
per-question
outputs, the
comparisons with
these rows are
descriptive, without
paired intervals.
NIFT-SQL reports the
real-question
complete@15 values
of these two rows
(Rao \& Gopinath,
2026, Table 9);
recall@5, recall@15,
and complete@5 come
from the same run
and are not reported
there.

\needspace{26\baselineskip}

\emph{Table 8.
BEAVER, real
questions (121) and
a sample of
synthetic questions
(1,000). ``BM25
rerank of pool''
ranks the 100
candidates by BM25
score. Rows trained
on BEAVER come from
the NIFT-SQL
confirmation run;
their complete@15
values are published
in Rao \& Gopinath
(2026, Table 9).}

\begin{longtable}[]{@{}
  >{\raggedright\arraybackslash}p{(\columnwidth - 12\tabcolsep) * \real{0.1639}}
  >{\raggedright\arraybackslash}p{(\columnwidth - 12\tabcolsep) * \real{0.2623}}
  >{\raggedright\arraybackslash}p{(\columnwidth - 12\tabcolsep) * \real{0.1475}}
  >{\raggedright\arraybackslash}p{(\columnwidth - 12\tabcolsep) * \real{0.0984}}
  >{\raggedright\arraybackslash}p{(\columnwidth - 12\tabcolsep) * \real{0.1311}}
  >{\raggedright\arraybackslash}p{(\columnwidth - 12\tabcolsep) * \real{0.0984}}
  >{\raggedright\arraybackslash}p{(\columnwidth - 12\tabcolsep) * \real{0.0984}}@{}}
\toprule\noalign{}
\begin{minipage}[b]{\linewidth}\raggedright
Questions
\end{minipage} &
\begin{minipage}[b]{\linewidth}\raggedright
Method
\end{minipage} &
\begin{minipage}[b]{\linewidth}\raggedright
Trained on BEAVER
\end{minipage} &
\begin{minipage}[b]{\linewidth}\raggedright
R@5
\end{minipage} &
\begin{minipage}[b]{\linewidth}\raggedright
R@15 {[}95\% CI{]}
\end{minipage} &
\begin{minipage}[b]{\linewidth}\raggedright
C@5
\end{minipage} &
\begin{minipage}[b]{\linewidth}\raggedright
C@15
\end{minipage} \\
\midrule\noalign{}
\endhead
\bottomrule\noalign{}
\endlastfoot
121 real & BM25
rerank of pool & no
& 48.9 & 70.4 & 16.5
& 29.8 \\
& pool & no & 55.3 &
78.8 & 20.7 &
39.7 \\
& FERN-schema & no &
57.0 & 82.1 {[}78.3,
85.5{]} & 22.3 &
49.6 \\
& NIFT-SQL & yes &
56.6 & 80.8 & 24.0 &
54.5 \\
& retriever
fine-tuned on BEAVER
(label-matched) &
yes & 53.0 & 84.5 &
22.3 & 62.0 \\
1,000 synthetic &
BM25 rerank of pool
& no & 53.9 & 83.9 &
13.7 & 53.3 \\
& pool & no & 54.6 &
88.0 & 16.5 &
61.0 \\
& FERN-schema & no &
61.7 & 88.9 {[}88.0,
89.9{]} & 22.5 &
63.3 \\
\end{longtable}

On the real
questions, FERN
improves on its pool
by 3.3 points of
recall@15 and 9.9
points of
complete@15. Its
recall@15 lies
between the two
BEAVER-trained
retrievers (80.8 and
84.5); its
complete@15 is 5.0
and 12.4 points
lower (6 and 15
questions). With
almost four gold
tables per question,
average recall can
look strong while a
query still misses a
required table, so
complete@15 is the
more relevant
number. With 121
questions, these
differences are
descriptive: a
complete@15 near
50\% has a 95\%
interval of about
plus or minus nine
points, and a
5.0-point difference
is six questions.
Both target-trained
baselines recovered
more complete table
sets on these
questions.

\needspace{8\baselineskip}

\hypertarget{where-the-signal-comes-from}{%
\subsubsection{6.3
Where the signal
comes
from}\label{where-the-signal-comes-from}}

We split the 120
features into
retrieval signals
(the score, agree,
and db families, 53
features) and the
other families (67),
with the same
hyperparameters. We
fitted each ranker
on all of BIRD train
and evaluated it on
BIRD dev;
separately, we ran
five-fold
cross-validation
grouped by database
within BIRD train
(all 9,394
questions). The PCA
projection of the
emb family, fitted
once on all of BIRD
train, is shared
across folds; it is
unsupervised but
sees the held-out
folds' unlabeled
rows, and the
retrieval-only
ranker does not use
it, so the
cross-validation is
a development
diagnostic rather
than a fully
held-out estimate.
We also remove one
family at a time
(BIRD dev). The
split is not
perfectly clean:
some of the other
features use the
pool's top
candidates as
reference points
(coverage beyond the
top table, join
distance to it, the
entropy of pool
scores).

\needspace{20\baselineskip}

\emph{Table 9.
Feature groups in
the frozen
configuration,
FERN-ev inputs. CV:
database-grouped
cross-validation
within BIRD train
with a shared
unsupervised PCA
(development
diagnostic). The
last two rows give
all features minus
the subset, paired
over questions, with
95\% bootstrap
intervals;
differences are
computed from
unrounded
per-question
values.}

\begin{longtable}[]{@{}
  >{\raggedright\arraybackslash}p{(\columnwidth - 16\tabcolsep) * \real{0.2373}}
  >{\raggedright\arraybackslash}p{(\columnwidth - 16\tabcolsep) * \real{0.0847}}
  >{\raggedright\arraybackslash}p{(\columnwidth - 16\tabcolsep) * \real{0.1186}}
  >{\raggedright\arraybackslash}p{(\columnwidth - 16\tabcolsep) * \real{0.0847}}
  >{\raggedright\arraybackslash}p{(\columnwidth - 16\tabcolsep) * \real{0.1017}}
  >{\raggedright\arraybackslash}p{(\columnwidth - 16\tabcolsep) * \real{0.0847}}
  >{\raggedright\arraybackslash}p{(\columnwidth - 16\tabcolsep) * \real{0.1017}}
  >{\raggedright\arraybackslash}p{(\columnwidth - 16\tabcolsep) * \real{0.1017}}
  >{\raggedright\arraybackslash}p{(\columnwidth - 16\tabcolsep) * \real{0.0847}}@{}}
\toprule\noalign{}
\begin{minipage}[b]{\linewidth}\raggedright
Features
\end{minipage} &
\begin{minipage}[b]{\linewidth}\raggedright
CV R@5
\end{minipage} &
\begin{minipage}[b]{\linewidth}\raggedright
CV C@5
\end{minipage} &
\begin{minipage}[b]{\linewidth}\raggedright
CV R@15
\end{minipage} &
\begin{minipage}[b]{\linewidth}\raggedright
CV C@15
\end{minipage} &
\begin{minipage}[b]{\linewidth}\raggedright
dev R@5
\end{minipage} &
\begin{minipage}[b]{\linewidth}\raggedright
dev C@5
\end{minipage} &
\begin{minipage}[b]{\linewidth}\raggedright
dev R@15
\end{minipage} &
\begin{minipage}[b]{\linewidth}\raggedright
dev C@15
\end{minipage} \\
\midrule\noalign{}
\endhead
\bottomrule\noalign{}
\endlastfoot
retrieval signals
only (53) & 78.6 &
70.0 & 90.7 & 86.8 &
91.9 & 86.2 & 98.2 &
96.6 \\
other families only
(67) & 86.4 & 77.1 &
94.6 & 89.9 & 95.4 &
91.9 & 98.5 &
97.1 \\
all (120) & 89.2 &
82.1 & 95.8 & 92.8 &
96.6 & 93.9 & 98.9 &
97.7 \\
all minus
retrieval-only & &
+12.1 {[}11.4,
12.9{]} & & +6.0
{[}5.5, 6.5{]} & &
+7.7 {[}6.3, 9.1{]}
& +0.6 {[}0.3,
1.0{]} & \\
all minus other-only
& & +5.0 {[}4.5,
5.6{]} & & +2.9
{[}2.5, 3.3{]} & &
+2.0 {[}1.0, 2.9{]}
& +0.4 {[}0.1,
0.7{]} & \\
\end{longtable}

\textbf{Both groups
reach high recall at
fifteen; combining
them matters at
five.} On BIRD dev,
a ranker over
retrieval signals
alone reaches 98.2
recall at fifteen
and one over the
other families 98.5;
all features reach
98.9 (differences
+0.6 and +0.4
computed before
rounding, both
intervals above
zero). The large
effect is at five
tables: combining
the groups raises
complete@5 by 7.7
points over
retrieval signals
alone and by 2.0
over the other
families. The
database-grouped
cross-validation
within BIRD train, a
development
diagnostic, shows
the same ordering
(12.1 and 5.0
points). On BIRD
with evidence, the
non-retrieval
features pay off
most when only a few
tables can be passed
to a generator; we
did not run these
feature-group
comparisons on the
transfer datasets.

\needspace{19\baselineskip}

\emph{Table 10.
Removing one family
(BIRD dev; all
features minus the
reduced set, 95\%
paired bootstrap,
computed from
unrounded values;
twelve exploratory
comparisons, not
corrected for
multiplicity).}

\begin{longtable}[]{@{}
  >{\raggedright\arraybackslash}p{(\columnwidth - 4\tabcolsep) * \real{0.3115}}
  >{\raggedright\arraybackslash}p{(\columnwidth - 4\tabcolsep) * \real{0.3770}}
  >{\raggedright\arraybackslash}p{(\columnwidth - 4\tabcolsep) * \real{0.3115}}@{}}
\toprule\noalign{}
\begin{minipage}[b]{\linewidth}\raggedright
Family removed
\end{minipage} &
\begin{minipage}[b]{\linewidth}\raggedright
dev C@5
\end{minipage} &
\begin{minipage}[b]{\linewidth}\raggedright
dev R@15
\end{minipage} \\
\midrule\noalign{}
\endhead
\bottomrule\noalign{}
\endlastfoot
db (14) & +0.8
{[}0.1, 1.6{]} &
+0.2 {[}0.1,
0.4{]} \\
graph (7) & +0.8
{[}0.1, 1.6{]} &
-0.1 {[}-0.2,
0.1{]} \\
val (6) & +0.7
{[}0.0, 1.4{]} &
+0.1 {[}-0.1,
0.4{]} \\
lex (8) & +0.3
{[}-0.4, 1.0{]} &
+0.1 {[}-0.1,
0.3{]} \\
score, agree, cov,
ev, type, meta, q,
emb & between -0.3
and +0.1, all
intervals include
zero & between -0.1
and 0.0 \\
\end{longtable}

\textbf{Few single
families have a
measurable effect.}
Removing database or
graph features
lowered BIRD dev
complete@5 by 0.8
points; removing
value features
lowered it by 0.7,
with the interval
touching zero. For
the other nine
families, the
exploratory
complete@5 intervals
included zero. These
ablations do not
establish whether
that reflects
redundancy between
families (the score
family, for example,
overlaps with the
agreement and
database families,
which are built from
the same scores) or
whether the family
contains little
information. In the
fitted ranker, the
most important
features by gain are
the pool score's gap
to the best
candidate (49.5\% of
the total gain), the
pool rank (13.3\%)
and the best pool
score in the table's
database (3.5\%).

Appendix B reports
the development
studies that led to
this configuration,
including a
fine-tuned decision
model that added
little.

\needspace{8\baselineskip}

\hypertarget{cost}{%
\subsubsection{6.4
Cost}\label{cost}}

\needspace{21\baselineskip}

\emph{Table 11.
Measured retrieval
latency per
question, FERN-ev,
batch size 1, one
NVIDIA A100 40GB,
100 BIRD dev
questions after 5
warm-up questions;
indexes and card
embeddings
precomputed.}

\begin{longtable}[]{@{}
  >{\raggedright\arraybackslash}p{(\columnwidth - 4\tabcolsep) * \real{0.3333}}
  >{\raggedright\arraybackslash}p{(\columnwidth - 4\tabcolsep) * \real{0.2667}}
  >{\raggedright\arraybackslash}p{(\columnwidth - 4\tabcolsep) * \real{0.4000}}@{}}
\toprule\noalign{}
\begin{minipage}[b]{\linewidth}\raggedright
Stage
\end{minipage} &
\begin{minipage}[b]{\linewidth}\raggedright
Median (ms)
\end{minipage} &
\begin{minipage}[b]{\linewidth}\raggedright
90th percentile (ms)
\end{minipage} \\
\midrule\noalign{}
\endhead
\bottomrule\noalign{}
\endlastfoot
embed the question
with Qwen3 and BGE
(GPU) & 71.9 &
87.6 \\
BM25, two views
(CPU) & 6.6 &
10.5 \\
dense scores and
pool & 0.4 & 12.8 \\
features (CPU,
Python) & 122.7 &
146.0 \\
ranker (CPU) & 6.4 &
10.2 \\
total & 208.3 &
246.9 \\
\end{longtable}

Stage medians need
not add up to the
median total. FERN
makes no hosted
model calls and
sends no schema
outside the system.
Together, the two
embedders have under
one billion
parameters. Fitting
the three rankers
takes about three
and a half minutes
on a CPU; computing
features for the
9,394 training
questions takes 11
minutes with
evidence and 4
without (in bulk,
without the
per-question
overheads included
in Table 11). The
features are plain
Python and have not
been optimized. RASL
estimates 1.85 US
dollars per 100 BIRD
queries and 0.80 per
100 Spider queries
in its input-token
cost analysis, which
includes SQL
generation, with 3
cents per 100 in the
category it labels
retrieval (Eben et
al., 2025); it does
not report latency,
so we make no
latency comparison.

\needspace{8\baselineskip}

\hypertarget{discussion}{%
\subsection{7
Discussion}\label{discussion}}

\textbf{BIRD and
Spider.} On these
benchmarks, whose
table and column
names are mostly
readable English,
frozen retrievers
already place almost
every gold table in
the candidate pool.
The study does not
isolate readability
as the variable that
explains the
difference from
Fiben. What remains
is ordering the
candidates. At a
nominal budget of
five, FERN's recall
is 0.3 points below
RASL's published
value on Spider and
0.9 points below on
BIRD with evidence;
the comparison does
not isolate the
contribution of LLM
selection. We do not
know why RASL is
ahead on BIRD at
five tables; one
hypothesis is the
LLM's ability to
judge which of
several plausible
tables a question
needs, another is
RASL's use of BIRD's
column and value
descriptions.

\textbf{Evidence.}
BIRD's evidence
hints often name the
exact columns a
question needs. With
evidence, FERN's
recall at five is
7.3 points higher
(96.6 vs 89.3). The
7.3-point difference
compares two
separately fitted
variants that differ
in both the query
text and the
evidence-specific
features; it is not
decomposed between
them. For
deployments without
per-question
evidence hints, the
question-only
results are more
relevant, including
the 8.2-point gap at
five tables to
RASL's published
number.

\textbf{Opaque
schemas.} Fiben's
tables have
run-together
uppercase names (for
example,
\texttt{ACCOUNTSPAYABLE\allowbreak{}ANDACCRUED\allowbreak{}LIABILITIES}),
2.5 columns per
table, one database,
and almost four gold
tables per question.
Reranking the pool
by BM25 and the pool
itself both reach
only 39 recall at
fifteen; FERN
improves on its pool
by 15 points but
remains far below
RASL. The results
expose limitations
in candidate
retrieval and in
ranking under a
small output budget.
The fused pool is
weak: 67.0 recall at
fifty tables,
against 79.4 for
RASL's question-only
retriever and 90.6
for its pool using
the question plus
LLM-extracted
keywords (Eben et
al., 2025, Table 5),
and its average
recall at 100 is
capped at 82.8;
standalone BM25 over
the whole catalog
reaches only 7.0 at
fifteen.

Fiben has only 152
tables, so fifty is
a third of the
catalog. After
reranking its 100
candidates, FERN
reaches 80.7 recall
when returning fifty
tables, but only
54.2 when returning
fifteen, a drop of
26.5 points. RASL's
question-only
configuration
reports 79.4 recall
for its initial
fifty-table pool and
67.5 after LLM
selection with a
maximum output of
fifteen, a drop of
11.9 points. These
figures suggest that
retaining relevant
tables under a small
output budget is a
substantial
difficulty for FERN.
They do not isolate
the effect of LLM
selection, because
the systems use
different question
sets, embeddings,
and retrieval
representations;
RASL retrieves
individual schema
entities with Cohere
embeddings. Fiben
also differs from
BIRD and Spider in
several ways at
once: its names are
unsegmented, our
copy has no cell
values, it has a
single database
(which removes
discrimination
between databases
but preserves
within-database
ranking signals),
and its questions
need many tables. We
cannot tell which of
these matters.

\textbf{Enterprise
use.} On the 121
real dw questions,
with no BEAVER
question used for
fitting (although
BEAVER results
informed one design
decision; Section
5), FERN's recall at
fifteen fell between
the two
target-trained point
estimates, while its
complete@15 was
lower than both.
Both baselines were
trained on BEAVER's
synthetic questions
(Table 8); a
retriever fine-tuned
this way could enter
FERN as another
signal, which we
have not tested. We
see public-data
fitting and
target-catalog
supervision as
stages of a
deployment: a ranker
fitted on public
data first,
supervision on the
target catalog once
data exists.

\needspace{8\baselineskip}

\hypertarget{limitations}{%
\subsection{8
Limitations}\label{limitations}}

\begin{itemize}
\tightlist
\item
  \textbf{Published
  reference
  numbers.} We did
  not rerun RASL;
  its per-question
  results are
  unavailable, and
  its inputs differ
  (Table 2): it uses
  BIRD's
  descriptions, not
  cell values, and
  returns at most N
  tables.
  Differences from
  RASL carry no
  interval. The
  Fiben question
  sets differ.
\item
  \textbf{Development
  exposure.} BIRD
  dev and BEAVER
  informed two
  design decisions
  (Section 5);
  Spider and Fiben
  did not.
\item
  \textbf{Supervision.}
  The rankers are
  fitted on 9,394
  labeled BIRD
  questions, far
  more than RASL's
  200 calibration
  questions. FERN
  avoids
  target-domain
  fitting, not
  supervision.
  Pretraining of the
  frozen embedders
  on public
  benchmarks cannot
  be ruled out, as
  for any pretrained
  model.
\item
  \textbf{Evidence
  and values.}
  FERN's best BIRD
  numbers use
  evidence and cell
  values. Without
  evidence, recall
  at five is 8.2
  points below
  RASL's published
  number.
\item
  \textbf{Gold
  labels.}
  Gold-table
  extraction changes
  1.6\% of BIRD dev
  gold sets, the
  same order as the
  differences from
  RASL; RASL does
  not describe its
  extraction. BIRD's
  gold SQL itself
  contains
  annotation errors
  (Wretblad et al.,
  2024; Jin et al.,
  2026), which can
  affect both FERN's
  and RASL's
  reported recall.
\item
  \textbf{Candidate
  ceiling.} Gold
  tables outside the
  100 candidates
  cannot be
  recovered (average
  recall at 100 is
  99.4 on BIRD dev
  and 82.8 on
  Fiben).
\item
  \textbf{End-to-end
  scope.} We
  evaluate table
  retrieval only,
  not database
  routing or the SQL
  generated from the
  retrieved tables.
\item
  \textbf{Benchmarks.}
  We evaluate RASL's
  three benchmarks
  and one enterprise
  benchmark. Larger
  enterprise suites
  with thousands of
  tables need
  candidate-pool
  decisions of their
  own and are left
  to future work.
\end{itemize}

\needspace{8\baselineskip}

\hypertarget{conclusion}{%
\subsection{9
Conclusion}\label{conclusion}}

Against RASL's
published results on
pooled catalogs,
rankers fitted once
on BIRD over frozen
retrievers come
within one point of
its published recall
at five tables on
Spider and on BIRD
with evidence, at
about 0.2 seconds
per question with no
hosted model calls;
without evidence,
the gap on BIRD at
five tables is eight
points. On Spider,
the candidate pool
of frozen retrievers
(BM25, Qwen3 and
BGE) alone already
reaches RASL's
recall at fifteen,
and the ranker's
main contribution is
complete table sets
at small budgets. On
the real questions
of an enterprise
benchmark, the
rankers' recall at
fifteen falls
between that of two
retrievers trained
on it, while their
complete@15 is lower
than both. On Fiben,
their recall is
numerically close to
RASL's question-only
retrieval result at
fifty tables but
substantially lower
than RASL's selected
tables at fifteen;
both candidate
retrieval and
ranking under a
small budget remain
limitations. FERN
was competitive in
reported recall on
Spider and on BIRD
with evidence, which
makes it a strong
baseline on these
benchmarks; Fiben
shows where it is
not enough.

\needspace{8\baselineskip}

\hypertarget{data-and-code-availability}{%
\subsection{Data and
code
availability}\label{data-and-code-availability}}

The code, frozen
configuration,
catalog and question
manifests, fitted
rankers and
per-question results
will be released
upon journal
publication. BEAVER
requires accepting
its terms on Hugging
Face.

\needspace{8\baselineskip}

\hypertarget{references}{%
\subsection{References}\label{references}}

Bruch, S., Gai, S.,
\& Ingber, A.
(2023). An analysis
of fusion functions
for hybrid
retrieval. ACM
Transactions on
Information Systems,
42(1), Article 20.

Burges, C. J. C.
(2010). From RankNet
to LambdaRank to
LambdaMART: An
overview. Microsoft
Research Technical
Report
MSR-TR-2010-82.

Chen, P. B., Zhang,
Y., \& Roth, D.
(2024a). Is table
retrieval a solved
problem? Exploring
join-aware
multi-table
retrieval. In
Proceedings of ACL
2024.

Chen, P. B., Yang,
D., Li, W., Wenz,
F., Zhang, Y.,
Tatbul, N.,
Cafarella, M.,
Demiralp, Ç., \&
Stonebraker, M.
(2024b). BEAVER: An
enterprise benchmark
for text-to-SQL.
arXiv:2409.02038v3.

Cormack, G. V.,
Clarke, C. L. A., \&
Büttcher, S. (2009).
Reciprocal rank
fusion outperforms
Condorcet and
individual rank
learning methods. In
Proceedings of SIGIR
2009.

Eben, J., Ahmad, A.,
\& Lau, S. (2025).
RASL: Retrieval
augmented schema
linking for massive
database
text-to-SQL. KDD
2025 Workshop on
Structured Knowledge
for Large Language
Models (SKnowLLM).
arXiv:2507.23104.

Guo, J., Zhan, Z.,
Gao, Y., Xiao, Y.,
Lou, J.-G., Liu, T.,
\& Zhang, D. (2019).
Towards complex
text-to-SQL in
cross-domain
database with
intermediate
representation. In
Proceedings of ACL
2019.

Herzig, J., Müller,
T., Krichene, S., \&
Eisenschlos, J. M.
(2021). Open domain
question answering
over tables via
dense retrieval. In
Proceedings of NAACL
2021.

Jin, T., Choi, Y.,
Zhu, Y., \& Kang, D.
(2026). Pervasive
annotation errors
break text-to-SQL
benchmarks and
leaderboards.
arXiv:2601.08778v3.

Ke, G., Meng, Q.,
Finley, T., Wang,
T., Chen, W., Ma,
W., Ye, Q., \& Liu,
T.-Y. (2017).
LightGBM: A highly
efficient gradient
boosting decision
tree. In Advances in
Neural Information
Processing Systems
30, 3146 to 3154.

Kothyari, M.,
Dhingra, D.,
Sarawagi, S., \&
Chakrabarti, S.
(2023). CRUSH4SQL:
Collective retrieval
using schema
hallucination for
Text2SQL. In
Proceedings of EMNLP
2023.

Li, H., Zhang, J.,
Li, C., \& Chen, H.
(2023). RESDSQL:
Decoupling schema
linking and skeleton
parsing for
text-to-SQL. In
Proceedings of AAAI
2023.

Li, H., Zhang, J.,
Liu, H., Fan, J.,
Zhang, X., Zhu, J.,
Wei, R., Pan, H.,
Li, C., \& Chen, H.
(2024). CodeS:
Towards building
open-source language
models for
text-to-SQL.
Proceedings of the
ACM on Management of
Data (SIGMOD 2024).

Li, J., Hui, B., Qu,
G., Yang, J., Li,
B., Li, B., Wang,
B., Qin, B., Cao,
R., Geng, R., Huo,
N., Zhou, X., Ma,
C., Li, G., Chang,
K. C. C., Huang, F.,
Cheng, R., \& Li, Y.
(2023). Can LLM
already serve as a
database interface?
A BIg bench for
large-scale database
grounded
text-to-SQLs. In
Advances in Neural
Information
Processing Systems
36, Datasets and
Benchmarks Track.

Maamari, K.,
Abubaker, F.,
Jaroslawicz, D., \&
Mhedhbi, A. (2024).
The death of schema
linking? Text-to-SQL
in the age of
well-reasoned
language models.
arXiv:2408.07702.

Page, L., Brin, S.,
Motwani, R., \&
Winograd, T. (1999).
The PageRank
citation ranking:
Bringing order to
the web. Stanford
InfoLab Technical
Report.

Pourreza, M., \&
Rafiei, D. (2023).
DIN-SQL: Decomposed
in-context learning
of text-to-SQL with
self-correction. In
Advances in Neural
Information
Processing Systems
36.

Qin, T., \& Liu,
T.-Y. (2013).
Introducing LETOR
4.0 datasets.
arXiv:1306.2597.

Qin, Z., Yan, L.,
Zhuang, H., Tay, Y.,
Pasumarthi, R. K.,
Wang, X., Bendersky,
M., \& Najork, M.
(2021). Are neural
rankers still
outperformed by
gradient boosted
decision trees? In
Proceedings of ICLR
2021.

Rao, A., \&
Gopinath, N. (2026).
NIFT-SQL:
Customer-supervised
table scoring for
enterprise
text-to-SQL
retrieval. arXiv
preprint
arXiv:2611.12731.

Robertson, S., \&
Zaragoza, H. (2009).
The probabilistic
relevance framework:
BM25 and beyond.
Foundations and
Trends in
Information
Retrieval, 3(4), 333
to 389.

Sen, J., Lei, C.,
Quamar, A., Özcan,
F., Efthymiou, V.,
Dalmia, A., Stager,
G., Mittal, A.,
Saha, D., \&
Sankaranarayanan, K.
(2020). ATHENA++:
Natural language
querying for complex
nested SQL queries.
Proceedings of the
VLDB Endowment,
13(11), 2747 to
2759.

Singh, R., Boškov,
N., Drabeck, L.,
Gudal, A., \& Khan,
M. A. (2026).
CSR-RAG: An
efficient retrieval
system for
text-to-SQL on the
enterprise scale.
arXiv:2601.06564.

Soliman, H., Gupta,
V., Roth, D., \&
Gurevych, I. (2026).
CORE-T: COherent
REtrieval of Tables
for Text-to-SQL.
arXiv:2601.13111v3.

sqlglot (2024). SQL
parser and
transpiler.
\url{https://github.com/tobymao/sqlglot}.

Talaei, S.,
Pourreza, M., Chang,
Y.-C., Mirhoseini,
A., \& Saberi, A.
(2024). CHESS:
Contextual
harnessing for
efficient SQL
synthesis.
arXiv:2405.16755.

Wang, B., Shin, R.,
Liu, X., Polozov,
O., \& Richardson,
M. (2020). RAT-SQL:
Relation-aware
schema encoding and
linking for
text-to-SQL parsers.
In Proceedings of
ACL 2020.

Wang, T., Chen, X.,
Lin, H., Han, X.,
Sun, L., Wang, H.,
\& Zeng, Z. (2025).
DBCopilot: Natural
language querying
over massive
databases via schema
routing.
arXiv:2312.03463v3.

Wang, Y., Liu, P.,
\& Yang, X. (2025).
LinkAlign: Scalable
schema linking for
real-world
large-scale
multi-database
text-to-SQL. In
Proceedings of EMNLP
2025, 977 to 991.

Warner, B., Chaffin,
A., Clavié, B.,
Weller, O.,
Hallström, O.,
Taghadouini, S.,
Gallagher, A.,
Biswas, R., Ladhak,
F., Aarsen, T.,
Cooper, N., Adams,
G., Howard, J., \&
Poli, I. (2024).
Smarter, better,
faster, longer: A
modern bidirectional
encoder for fast,
memory efficient,
and long context
finetuning and
inference.
arXiv:2412.13663.

Wretblad, N.,
Riseby, F. G.,
Biswas, R., Ahmadi,
A., \& Holmström, O.
(2024).
Understanding the
effects of noise in
text-to-SQL: An
examination of the
BIRD-Bench
benchmark. In
Proceedings of ACL
2024 (Volume 2:
Short Papers), 356
to 369.

Xiao, S., Liu, Z.,
Zhang, P.,
Muennighoff, N.,
Lian, D., \& Nie,
J.-Y. (2024).
C-Pack: Packed
resources for
general Chinese
embeddings. In
Proceedings of SIGIR
2024.

Yu, T., Zhang, R.,
Yang, K., Yasunaga,
M., Wang, D., Li,
Z., Ma, J., Li, I.,
Yao, Q., Roman, S.,
Zhang, Z., \& Radev,
D. (2018). Spider: A
large-scale
human-labeled
dataset for complex
and cross-domain
semantic parsing and
text-to-SQL task. In
Proceedings of EMNLP
2018.

Zhang, S., \& Balog,
K. (2018). Ad hoc
table retrieval
using semantic
similarity. In
Proceedings of The
Web Conference (WWW)
2018.

Zhang, Y., Li, M.,
Long, D., Zhang, X.,
Lin, H., Yang, B.,
Xie, P., Yang, A.,
Liu, D., Lin, J.,
Huang, F., \& Zhou,
J. (2025). Qwen3
Embedding: Advancing
text embedding and
reranking through
foundation models.
arXiv:2506.05176.

\needspace{8\baselineskip}

\clearpage

\hypertarget{appendix-a-design-selection-on-bird-train}{%
\subsection{Appendix
A: Design selection
on BIRD
train}\label{appendix-a-design-selection-on-bird-train}}

\needspace{25\baselineskip}

\emph{Table A1.
Candidate pool
options, five-fold
cross-validation
(folds by database)
on a 3,000-question
subset of BIRD
train. Top 100
tables by reciprocal
rank fusion. ``All
six views'' includes
the best-column
signal later
removed.}

\begin{longtable}[]{@{}
  >{\raggedright\arraybackslash}p{(\columnwidth - 4\tabcolsep) * \real{0.4333}}
  >{\raggedright\arraybackslash}p{(\columnwidth - 4\tabcolsep) * \real{0.2833}}
  >{\raggedright\arraybackslash}p{(\columnwidth - 4\tabcolsep) * \real{0.2833}}@{}}
\toprule\noalign{}
\begin{minipage}[b]{\linewidth}\raggedright
Signals fused
\end{minipage} &
\begin{minipage}[b]{\linewidth}\raggedright
R@100
\end{minipage} &
\begin{minipage}[b]{\linewidth}\raggedright
C@100
\end{minipage} \\
\midrule\noalign{}
\endhead
\bottomrule\noalign{}
\endlastfoot
BM25 & 93.7 &
87.2 \\
Qwen3 & 98.8 &
97.5 \\
BGE & 98.4 & 96.8 \\
BM25 + Qwen3 & 98.9
& 97.7 \\
BM25 + BGE & 98.4 &
96.5 \\
Qwen3 + BGE
(selected by the
rule) & 99.3 &
98.4 \\
BM25 + Qwen3 + BGE
(used; Section 5,
exception 1) & 99.2
& 98.3 \\
all six views & 99.3
& 98.4 \\
\end{longtable}

\needspace{23\baselineskip}

\emph{Table A2.
Removing one signal
from the ranker,
same
cross-validation
(complete@5 /
complete@15). We
also checked
removing every
``drop'' signal
together; we dropped
only one signal.}

\begin{longtable}[]{@{}
  >{\raggedright\arraybackslash}p{(\columnwidth - 6\tabcolsep) * \real{0.3167}}
  >{\raggedright\arraybackslash}p{(\columnwidth - 6\tabcolsep) * \real{0.1833}}
  >{\raggedright\arraybackslash}p{(\columnwidth - 6\tabcolsep) * \real{0.1833}}
  >{\raggedright\arraybackslash}p{(\columnwidth - 6\tabcolsep) * \real{0.3167}}@{}}
\toprule\noalign{}
\begin{minipage}[b]{\linewidth}\raggedright
Ranker
\end{minipage} &
\begin{minipage}[b]{\linewidth}\raggedright
C@5
\end{minipage} &
\begin{minipage}[b]{\linewidth}\raggedright
C@15
\end{minipage} &
\begin{minipage}[b]{\linewidth}\raggedright
Decision
\end{minipage} \\
\midrule\noalign{}
\endhead
\bottomrule\noalign{}
\endlastfoot
all signals & 82.6 &
92.7 & \\
without BM25, full
card & 82.1 & 92.9 &
keep \\
without BM25, name
card & 81.8 & 93.1 &
keep \\
without Qwen3, full
card & 81.9 & 92.5 &
keep \\
without Qwen3, name
card & 81.6 & 92.8 &
keep \\
without Qwen3, best
column & 82.4 & 92.8
& drop \\
without BGE & 82.2 &
92.7 & keep \\
\end{longtable}

The selected and
used pools in Table
A1 differ by 0.1
points in both
metrics, and in
Table A2 each
single-signal
removal changes
complete@5 and
complete@15 by at
most 1.0 point
relative to the
all-signals ranker;
both are close to
cross-validation
noise with 3,000
questions, and their
value is that a
stated rule chose
each component.

\needspace{8\baselineskip}

\hypertarget{appendix-b-development-studies-history}{%
\subsection{Appendix
B: Development
studies
(history)}\label{appendix-b-development-studies-history}}

These studies were
run during
development on BIRD
dev, in a
configuration that
differs from the
final one: 139
features (the final
families plus a
fine-tuned decision
model and the
best-column signal),
a BM25 + Qwen3 pool,
rankers fitted on
3,000 BIRD training
questions with
out-of-fold
decision-model
scores, and, where
stated, name-regex
labels. They
motivated the
design; they do not
test it.

\needspace{23\baselineskip}

\emph{Table B1.
Development results
on BIRD dev with
evidence.
``Stacked'' is a
stacking ensemble of
a random forest (200
trees), a LightGBM
classifier (500
trees), and a
logistic regression,
combined by logistic
regression with
three-fold
cross-fitting.}

\begin{longtable}[]{@{}
  >{\raggedright\arraybackslash}p{(\columnwidth - 10\tabcolsep) * \real{0.3115}}
  >{\raggedright\arraybackslash}p{(\columnwidth - 10\tabcolsep) * \real{0.1639}}
  >{\raggedright\arraybackslash}p{(\columnwidth - 10\tabcolsep) * \real{0.1803}}
  >{\raggedright\arraybackslash}p{(\columnwidth - 10\tabcolsep) * \real{0.1148}}
  >{\raggedright\arraybackslash}p{(\columnwidth - 10\tabcolsep) * \real{0.1148}}
  >{\raggedright\arraybackslash}p{(\columnwidth - 10\tabcolsep) * \real{0.1148}}@{}}
\toprule\noalign{}
\begin{minipage}[b]{\linewidth}\raggedright
Ranker
\end{minipage} &
\begin{minipage}[b]{\linewidth}\raggedright
Labels
\end{minipage} &
\begin{minipage}[b]{\linewidth}\raggedright
Features
\end{minipage} &
\begin{minipage}[b]{\linewidth}\raggedright
R@15
\end{minipage} &
\begin{minipage}[b]{\linewidth}\raggedright
C@5
\end{minipage} &
\begin{minipage}[b]{\linewidth}\raggedright
C@15
\end{minipage} \\
\midrule\noalign{}
\endhead
\bottomrule\noalign{}
\endlastfoot
hybrid search (BM25
+ Qwen3) & sqlglot &
& 94.2 & 69.4 &
88.5 \\
12 score features,
stacked & regex & 12
& 97.9 & 87.0 &
96.1 \\
all features with
the decision model,
stacked & regex &
139 & 98.7 & 93.0 &
97.5 \\
retrieval signals
only, lambdarank &
sqlglot & 58 & 98.1
& 86.4 & 96.5 \\
other features only,
lambdarank & sqlglot
& 66 & 98.1 & 91.5 &
96.3 \\
all features without
the decision model,
stacked & sqlglot &
124 & 98.8 & 93.0 &
97.6 \\
all features with
the decision model,
stacked & sqlglot &
139 & 98.8 & 93.7 &
97.8 \\
\end{longtable}

The 12 score
features were the
hybrid score and
rank, the dense and
BM25 scores, the
decision model's
score and rank, and
six database
features. The
decision model was
OpenDecider-nano, a
0.4B
ModernBERT-based
encoder (Warner et
al., 2024),
fine-tuned as a
yes/no table scorer
on 2,500 questions
per fold. Including
it changed recall@15
by +0.1 (95\% paired
interval -0.1 to
+0.2), complete@15
by +0.3 (0.0 to
+0.6), and
complete@5 by +0.8
(+0.2 to +1.4), all
computed from
unrounded
per-question values,
at 336 of 570 ms per
question; it was
left out of FERN
(Section 5,
exception 2). BIRD
is close to the
ceiling at fifteen
tables, and this
scorer was trained
on few questions, so
this does not show
that learned scorers
are unhelpful in
general; on BEAVER,
retrievers trained
on the target
catalog recovered
more complete table
sets than FERN
(Table 8).

\end{document}